\documentstyle[floats,prl,aps]{revtex}
\begin{document}
\renewcommand{\thefootnote}{\fnsymbol{footnote}}
\draft
\title{\large\bf 
  Quantum integrability and exact solution of the
		 supersymmetric $U$ model with boundary terms}

\author{Yao-Zhong Zhang \footnote {
                                   E-mail: yzz@maths.uq.edu.au}
             and 
        Huan-Qiang Zhou \footnote {Present address: Dept of
	         Physics, Chongqing University, Chongqing 630044, China.
                 E-mail: hqzhou@cqu.edu.cn}} 

\address{      Department of Mathematics,University of Queensland,
		     Brisbane, Qld 4072, Australia}

\maketitle

\vspace{10pt}

\begin{abstract}
The quantum integrability is established for the one-dimensional
supersymmetric $U$ model with  boundary terms by means of the quantum
inverse scattering method. 
The boundary supersymmetric $U$ chain is solved by using
the coordinate space Bethe ansatz technique and Bethe ansatz equations
are derived. This provides us with a basis for computing the finite 
size corrections to the low lying energies in the system.
\end{abstract}

\pacs {PACS numbers: 71.10Fd, 75.10.Lp}



\def\a{\alpha}
\def\b{\beta}
\def\d{\delta}
\def\e{\epsilon}
\def\g{\gamma}
\def\k{\kappa}
\def\l{\lambda}
\def\o{\omega}
\def\t{\theta}
\def\s{\sigma}
\def\D{\Delta}
\def\L{\Lambda}


\def\beq{\begin{equation}}
\def\eeq{\end{equation}}
\def\bea{\begin{eqnarray}}
\def\eea{\end{eqnarray}}
\def\ba{\begin{array}}
\def\ea{\end{array}}
\def\no{\nonumber}
\def\le{\langle}
\def\re{\rangle}
\def\lt{\left}
\def\rt{\right}

\newcommand{\reff}[1]{eq.~(\ref{#1})}

\vskip.3in

In recent years, there has been a considerable interest in exactly 
solvable lattice models with boundary fields and/or interactions
\cite{Skl88,Mez91,deV93}. One
class of such models are one-dimensional boundary strongly correlated electron
systems, which is of great importance due to
their promising role in theoretical condensed-matter physics and 
possibly in high-$T_c$ superconductivity \cite{Ess92}. 
Boundary conditions for such systems, which are compatible with
integrability in the bulk, are constructed  
from solutions of the (graded) reflection equations (called boundary
K-matrices) \cite{Skl88}.
Work in this direction has been done 
for the Hubbard-like 
models \cite{Sch85,Zhou96,Asa96,Shi97}
and for the supersymmetric $t$-$J$ model \cite{Gon94,Ess96,Bed96}.  

In this Letter, we study integrable open-boundary conditions for the
supersymmetric $U$ model of strongly correlated electrons introduced
in \cite{Bra94,Bra95} and extensively investigated in 
\cite{Bed95,Ram96,Pfa96}. 
We will present a boundary supersymmetric $U$ model and 
show that it can be derived
from the quantum inverse scattering method by modifying and generalizing
Sklyanin's arguments, thus establishing the quantum integrability of
the boundary model. In doing so, we encounter
the following complication: the Sklyanin's definition for
boundary Hamiltonian can not apply since the supertrace of the boundary
K-matrices of zero spectral parameter is equal to zero for the
current case. This is related to the fact that the supersymmetric $U$ model
has been constructed from the R-matrix associated with the
{\em typical} 4-dimensional irreducible representation of $gl(2|1)$.
Nevertheless we manage to solve this complication by introducing a new
definition for Hamiltonian.
We then solve the boundary supersymmetric $U$ model by the coordinate
space Bethe ansatz approach and derive the Bethe Ansatz equations. 

Let $c_{j,\s}^\dagger$ and $c_{j,\s}$ denote fermionic creation and
annihilation operators with spin $\s$ at
site $j$, which satisfy the anti-commutation relations given by
$\{c_{i,\s}^\dagger, c_{j,\tau}\}=\d_{ij}\d_{\s\tau}$, where 
$i,j=1,2,\cdots,L$ and $\s,\tau=\uparrow,\;\downarrow$. We consider the
following Hamiltonian with boundary terms
\beq
H=\sum _{j=1}^{L-1} H_{j,j+1}^Q + B_{lt} +B_{rt},\label{h}
\eeq
where $H^Q_{j,j+1}$ is the local Hamiltonian of the supersymmetric
$U$ model introduced in \cite{Bra95} 
\bea
H_{j,j+1}^Q&=&-\sum _{\sigma}(c^{\dagger}_{j\sigma}c_{j+1\sigma}+h.c.)
  \exp(-\frac {1}{2}\eta n_{j,-\sigma}-\frac {1}{2}
  \eta n_{j+1,-\sigma})\no\\
& &+\frac {U}{2}(n_{j\uparrow}n_{j\downarrow}
  +n_{j+1\uparrow}n_{j+1\downarrow})
  +t_p(c^{\dagger}_{j\uparrow}c^{\dagger}_{j\downarrow}c_{j+1\downarrow}
  c_{j+1\uparrow}+h.c.)+(n_{j}+n_{j+1}),\label{local-h}
\eea
and $B_{lt},~B_{rt}$ are boundary terms
\beq
B_{lt}=-\frac {2(U+2)}{U(2-\xi_-)}\lt(\frac {2}{\xi_-}
  n_{1\uparrow}n_{1\downarrow}
  -n_1\rt),~~~~~~
B_{rt}=-\frac {2(U+2)}{U(2-\xi_+)}\lt(\frac {2}{\xi_+}
  n_{L\uparrow}n_{L\downarrow}
  -n_L\rt).\label{boundary-terms}
\eeq
In the above equations, $n_{j\s}$ is the number density operator
$n_{j\s}=c_{j\s}^{\dagger}c_{j\s}$,
$n_j=n_{j\uparrow}+n_{j\downarrow}$ and $t_p=\frac {U}{2} =
\pm (1-\exp(-\eta))$; $\xi_{\pm}$ are some parameters describing
boundary effects.

Some remarks are order. As is seen from (\ref{boundary-terms}),
$B_{lt} ~(B_{rt})$ is an inhomogenous combination of two terms 
contributing to the left (right) boundary conditions. The physical
meaning of these terms in the context of strongly correlated electrons
are the following. The first term is nothing but a boundary
on-site Coulomb interaction and the second term is a boundary chemical
potential. 

We will establish the quantum integrability of the boundary
supersymmetric $U$ model (\ref{h}) by showing that it can be derived from
the quantum inverse scattering method. Let us recall that the
Hamiltonian of 
the supersymmetric $U$ model with the periodic boundary conditions
commutes with the transfer matrix, which is the supertrace of the
monodromy matrix $T(u)$,
\beq
T(u) = R_{0L}(u)\cdots R_{01}(u). \label{matrix-t}
\eeq
The explicit form of the quantum R-matrix 
$ R_{0j}(u)$ is given in \cite{Bra94}.
Here $u$ is the spectral parameter, 
and the subscript $0$ denotes the auxiliary superspace $V=C^{2,2}$.
It should be noted that the supertrace
is carried out for the auxiliary superspace $V$.
The elements of the supermatrix $T(u)$ are the generators
of two associative superalgebra ${\cal A}$ defined by the relations
\beq
R_{12}(u_1-u_2) \stackrel {1}{T}(u_1) \stackrel {2}{T}(u_2) =
   \stackrel {2}{T}(u_2) \stackrel {1}{T}(u_1)R_{12}(u_1-u_2),\label{rtt-ttr} 
\eeq
where $\stackrel {1}{X} \equiv  X \otimes 1,~
\stackrel {2}{X} \equiv  1 \otimes X$
for any supermatrix $ X \in End(V) $. For later use, we list some useful
properties enjoyed by the R-matrix:
(i) Unitarity:   $  R_{12}(u)R_{21}(-u) = 1$ and (ii)
 Crossing-unitarity:  $  R^{st_2}_{12}(-u+2)R^{st_1}_{21}(u) =
         \tilde {\rho }(u)$
with $\tilde \rho (u)$ being a scalar function,
$\tilde \rho (u) =u^2(2-u)^2/[(2+2\alpha-u)^2(2\alpha+u)^2]$.
Throughout this letter,
$\alpha = \frac {2}{U}$.

In order to describe integrable electronic models on  open
chains, we introduce an associative
superalgebras ${\cal T}_-$  and ${\cal T}_+$ defined by the R-matrix
$R(u_1-u_2)$ and the relations
\beq
R_{12}(u_1-u_2)\stackrel {1}{\cal T}_-(u_1) R_{21}(u_1+u_2)
  \stackrel {2}{\cal T}_-(u_2)
=  \stackrel {2}{\cal T}_-(u_2) R_{12}(u_1+u_2)
  \stackrel {1}{\cal T}_-(u_1) R_{21}(u_1-u_2),  \label{reflection1}
\eeq
\bea
&&R_{21}^{st_1 ist_2}(-u_1+u_2)\stackrel {1}{\cal T}_+^{st_1}
  (u_1) R_{12}(-u_1-u_2+2)
  \stackrel {2}{\cal T}_+^{ist_2}(u_2)\no\\
&&~~~~~~~~~~~~~~~=\stackrel {2}{\cal T}_+^{ist_2}(u_2) R_{21}(-u_1-u_2+2)
  \stackrel {1}{\cal T}_+^{st_1}(u_1) R_{12}^{st_1 ist_2}(-u_1+u_2)
  \label{reflection2}
\eea
respectively. Here the supertransposition $st_{\mu}~(\mu =1,2)$ 
is only carried out in the
$\mu$-th factor superspace of $V \otimes V$, whereas $ist_{\mu}$ denotes
the inverse operation of  $st_{\mu}$. By modifying Sklyanin's 
arguments \cite{Skl88}, one
may show that the quantities $\tau(u)$ given by
$\tau (u) = str ({\cal T}_+(u){\cal T}_-(u))$
constitute a commutative family, i.e.,
        $[\tau (u_1),\tau (u_2)] = 0$. 

One can obtain a class of realizations of the superalgebras ${\cal T}_+$  and
${\cal T}_-$  by choosing  ${\cal T}_{\pm}(u)$ to be the form
\beq
{\cal T}_-(u) = T_-(u) \tilde {\cal T}_-(u) T^{-1}_-(-u),~~~~~~ 
{\cal T}^{st}_+(u) = T^{st}_+(u) \tilde {\cal T}^{st}_+(u) 
  \lt(T^{-1}_+(-u)\rt)^{st}\label{t-,t+} 
\eeq
with
\beq
T_-(u) = R_{0M}(u) \cdots R_{01}(u),~~~~
T_+(u) = R_{0L}(u) \cdots R_{0,M+1}(u),~~~~ 
\tilde {\cal T}_{\pm}(u) = K_{\pm}(u),
\eeq
where $K_{\pm}(u)$, called boundary K-matrices, 
are representations of  ${\cal T}_{\pm}  $ in Grassmann
algebra.

We now solve (\ref{reflection1}) and (\ref{reflection2}) 
for $K_+(u)$ and $K_-(u)$.
For simplicity, let us restrict ourselves to the diagonal case. Then, one may 
check that the matrix $K_-(u)$ given by
\beq
K_-(u)=  \frac {1}{ \xi_-(2-\xi_-)}\;{\rm diag}\, \lt(
A_-(u), B_-(u), B_-(u), C_-(u)\rt),\label{k-}
\eeq
where
$A_-(u)=(\xi_-+u)(2-\xi_--u),~
B_-(u)=(\xi_--u)(2-\xi_--u)$ and
$C_-(u)=(\xi_--u)(2-\xi_-+u)$,
satisfies (\ref{reflection1}).
The matrix $K_+(u)$ can be obtained from the isomorphism of the
superalgebras  ${\cal T}_-  $ and ${\cal T}_+  $. Indeed, given a solution
${\cal T}_- $ of (\ref{reflection1}), then ${\cal T}_+(u)$ defined by
\beq
{\cal T}_+^{st}(u) =  {\cal T}_-(-u+1)\label{t+t-}
\eeq
is a solution of (\ref{reflection2}). 
The proof follows from some algebraic computations upon
substituting (\ref{t+t-}) into  
(\ref{reflection2}) and making use
of the properties of the R-matrix .
Therefore, one may choose the boundary matrix $K_+(u)$ as 
\beq
K_+(u)={\rm diag}\, \lt(A_+(u), B_+(u), B_+(u), C_+(u)\rt)\label{k+}
\eeq
with
$A_+(u)=(2-2\alpha-\xi_+-u)(2\alpha+\xi_++u),~
B_+(u)=(-2\alpha-\xi_++u)(2\alpha+\xi_++u)$ and
$C_+(u)=(-2\alpha-\xi_++u)(2+2\alpha+\xi_+-u)$.

Now it can be shown  that 
Hamiltonian (\ref{h}) is related to the transfer matrix
$\tau (u)$ (up to an unimportant additive constant)
\bea
H&=&-\frac{2(U+2)}{U}H^R,\no\\
H^R&=&\frac {\tau'' (0)}{4(V+2W)}=
  \sum _{j=1}^{L-1} H^R_{j,j+1} + \frac {1}{2} \stackrel {1}{K'}_-(0)
+\frac {1}{2(V+2W)}\lt[str_0\lt(\stackrel {0}{K}_+(0)G_{L0}\rt)\rt.\no\\
& &\lt.+2\,str_0\lt(\stackrel {0}{K'}_+(0)H_{L0}^R\rt)+
  str_0\lt(\stackrel {0}{K}_+(0)\lt(H^R_{L0}\rt)^2\rt)\rt],\label{derived-h}
\eea
where 
\bea
V&=&str_0 K'_+(0),
~~~~~~~~W=str_0 \lt(\stackrel {0}{K}_+(0) H_{L0}^R\rt),\no\\
H^R_{i,j}&=&P_{i,j}R'_{i,j}(0),
~~~~~~~~G_{i,j}=P_{i,j}R''_{i,j}(0).
\eea
Here $P_{i,j}$ denotes the graded permutation operator acting on the $i$-th
and $j$-th quantum spaces. (\ref{derived-h}) implies that the boundary
supersymmetric $U$ model admits
an infinite number
of conserved currents which are in involution with each other, thus
assuring its integrability. It should be emphasized that 
Hamiltonian (\ref{h}) appears as the second derivative of the transfer matrix
$\tau (u)$ with respect to the spectral parameter $u$ at $u=0$. This
is due to the fact that the supertrace of $K_+(0)$ equals to zero.
As we mentioned before, the reason for the zero supertrace of $K_+(0)$
is related to the fact that the quantum space is the
4-dimensional {\em typical} irreducible representation of $gl(2|1)$. 
A similar situation also occurs in the Hubbard-like models \cite{Zhou96}.

Having established the quantum integrability of the boundary 
supersymmetric $U$ model, we now  solve it by using
the coordinate space Bethe ansatz method. 
Let us assume that the eigenfunction of Hamiltonian (\ref{h}) has the
form
\bea
| \Psi \rangle& =&\sum _{\{(x_j,\s _j)\}}\Psi _{\s_1,\cdots,\s_N}
  (x_1,\cdots, x_N)\,c^{\dagger}
  _{x_1\s_1}\cdots c^{\dagger}_{x_N\s_N} | 0 \rangle,\no\\
\Psi_{\s1,\cdots,\s N}(x_1,\cdots,x_N)
&=&\sum _P \e _P A_{\s_{Q1},\cdots,\s_{QN}}(k_{PQ1},\cdots,k_{PQN})
  \exp (i\sum ^N_{j=1} k_{P_j}x_j),
\eea
where the summation is taken over all permutations and negations of
$k_1,\cdots,k_N,$ and $Q$ is the permutation of the $N$ particles such that
$1\leq   x_{Q1}\leq   \cdots  \leq  x_{QN}\leq   L$.
The symbol $\e_P$ is a sign factor $\pm1$ and changes its sign
under each 'mutation'. Substituting the wavefunction into the
eigenvalue equation $ H| \Psi  \rangle = E | \Psi \rangle $,
one gets
\bea
A_{\cdots,\s_j,\s_i,\cdots}(\cdots,k_j,k_i,\cdots)&=&S_{ij}(k_i,k_j)
    A_{\cdots,\s_i,\s_j,\cdots}(\cdots,k_i,k_j,\cdots),\no\\
A_{\s_i,\cdots}(-k_j,\cdots)&=&s^L(k_j;p_{1\s_i})A_{\s_i,\cdots}
    (k_j,\cdots),\no\\
A_{\cdots,\s_i}(\cdots,-k_j)&=&s^R(k_j;p_{L\s_i})A_{\cdots,\s_i}(\cdots,k_j)
\eea
with  $S_{ij}(k_i,k_j)$ being the two particle scattering matrix  and
$s^L,~s^R$ the  boundary scattering matrices,
\bea
S_{ij}(k_i,k_j)&=&\frac {\theta (k_i)-\theta (k_j)+ic{\cal P}_{ij}}
 {\theta (k_i)-\theta (k_j)+ic},\no\\
s^L(k_j;p_{1\s_i})&=&\frac {1-p_{1\s_i}e^{ik_j}}
{1-p_{1\s_i}e^{-ik_j}},\no\\
s^R(k_j;p_{L\s_i})&=&\frac {1-p_{L\s_i}e^{-ik_j}}
{1-p_{L\s_i}e^{ik_j}}e^{2ik_j(L+1)}
\eea
where $p_{1\s}\equiv p_1=-1+\frac {2(U+2)}{U} \frac {1}{2-\xi_-},~~
p_{L\s}\equiv p_L=-1+\frac {2(U+2)}{U} \frac {1}{2-\xi_+}$ and
$c=e^\eta-1$;
${\cal P}_{ij}$ is a spin permutation operator,and the charge rapidities 
$\theta(k_j)$ are related to the single-particle
quasi-momenta $k_j$ by $\theta (k)=\frac {1}{2} \tan (\frac {k}{2})$
\cite{Bed95}. Then, the diagonalization of Hamiltonian (\ref{h}) reduces 
to solving  the following matrix  eigenvalue equation
\beq
T_j\;t=t,~~~~~~~j=1,\cdots,N,
\eeq
where $t$ denotes an eigenvector on the space of the spin variables
and $T_j$ takes the form
\beq
T_j=S_j^-(k_j)s^L(-k_j;p_{1\s_j})R^-_j(k_j)R^+_j(k_j)
    s^R(k_j;p_{L\s_j})S^+_j(k_j)
\eeq
with
\bea
S_j^+(k_j)&=&S_{j,N}(k_j,k_N) \cdots S_{j,j+1}(k_j,k_{j+1}),\no\\
S^-_j(k_j)&=&S_{j,j-1}(k_j,k_{j-1})\cdots S_{j,1}(k_j,k_1),\no\\
R^-_j(k_j)&=&S_{1,j}(k_1,-k_j)\cdots S_{j-1,j}(k_{j-1},-k_j),\no\\
R^+_j(k_j)&=&S_{j+1,j}(k_{j+1},-k_j)\cdots S_{N,j}(k_N,-k_j).
\eea
This problem may be solved using the algebraic Bethe Ansatz method.
The Bethe ansatz equations are 
\bea
e^{ik_j2(L+1)}\zeta(k_j;p_1)\zeta(k_j;p_L)
&=&\prod ^M_{\beta =1}\frac {\theta _j-\lambda_\beta +i\frac {c}{2}}
  {\theta _j-\lambda_\beta -i\frac {c}{2}}
  \cdot\frac {\theta _j+\lambda_\beta +i\frac {c}{2}}
  {\theta _j+\lambda_\beta -i\frac {c}{2}},\no\\
\prod ^N_{j =1}\frac {\lambda _{\alpha}-\theta _j +i\frac {c}{2}}
  {\lambda _{\alpha}-\theta_j -i\frac {c}{2}}
  \cdot\frac {\lambda_{\alpha}+\theta_j +i\frac {c}{2}}
  {\lambda_{\alpha}+\theta _j -i\frac {c}{2}}
&=&
  \prod ^M_{\stackrel {\beta =1}{\beta \neq \alpha}}\frac
  {\lambda _{\alpha}-\lambda _{\beta} +ic}
  {\lambda _{\alpha}-\lambda_{\beta} -ic}
  \cdot\frac {\lambda_{\alpha}+\lambda_{\beta} +ic}
  {\lambda_{\alpha}+\lambda _{\beta} -ic},
\eea
where $\t_j\equiv \t(k_j)$ and $\zeta (k;p)= (1-pe^{-ik})/(1-pe^{ik})$.
The energy eigenvalue $E$ of the model is given by
$E=-2\sum ^N_{j=1}\cos k_j$, where we have dropped an additive constant.

In conclusion, we have studied integrable open-boundary conditions for the
supersymmetric $U$ model. Its quantum integrability follows from the fact
that the Hamiltonian of the model on the open chain may be embbeded into
a one-parameter family of commuting transfer matrices. Moreover, the Bethe
Ansatz equations are derived with the use of the coordinate space Bethe ansatz
approach. This provides us with a basis for computing the finite size
corrections to the low lying energies in the system, which in turn allow
us to use the boundary conformal field theory technique to study
the critical properties of the boundary.
The details will be treated in a separate publication.

\vskip.3in
This work is supported by Australian Research Council, University of
Queensland New Staff Research Grant and Enabling Research Grant. H.-Q.Z
would like to thank Department of Mathematics of UQ for hospitality.



\begin{thebibliography}{99}
\bibitem{Skl88} E.K. Sklyanin, J. Phys. {\bf A:} Math.Gen. {\bf 21} (1988)
   2375.
\bibitem{Mez91} L. Mezincescu, R. Nepomechie, J. Phys. {\bf A:} Math.
   Gen. {\bf 24} (1991) L17.
\bibitem{deV93} H.J. de Vega, A. Gonz\'alez-Ruiz, J. Phys. {\bf A:}
   Math. Gen. {\bf 26} (1993) L519.
\bibitem {Ess92} F.H.L. Essler, V.E. Korepin, K. Schoutens,
   Phys. Rev. Lett. {\bf 68} (1992) 2960; {\bf 70} (1993) 73. 
\bibitem{Sch85} H.J. Schulz, J. Phys. {\bf C:} Solid State Phys. {\bf 18} 
   (1985) 581.
\bibitem{Zhou96} H.-Q. Zhou, Phys. Rev. {\bf B54} (1996) 41; ibid
   {\bf B53} (1996) 5089.
\bibitem{Asa96} H. Asakawa, M. Suzuki, J. Phys. {\bf A:} Math. Gen. 
   {\bf 29} (1996) 225.
\bibitem{Shi97} M. Shiroishi, M. Wadati, J. Phys. Soc. Jpn. {\bf 66} 
   (1997) 1.
\bibitem{Gon94} A. Gonz\'alez-Ruiz, Nucl. Phys. {\bf B424} (1994) 553.
\bibitem{Ess96} F.H. Essler, J. Phys. {\bf A:} Math. Gen. {\bf 29}
   (1996) 6183.
\bibitem{Bed96} G. Bed\"urftig, F.H. Essler, H. Frahm, Phys. Rev. Lett.
   {\bf 77} (1996) 5098.
\bibitem {Bra94}  A.J. Bracken, G.W. Delius, M.D. Gould, Y.-Z. Zhang,  
   J. Phys. {\bf A:} Math. Gen. {\bf 27} (1994) 6551.
\bibitem{Bra95} A.J. Bracken, M.D. Gould, J.R. Links, Y.-Z. Zhang,
   Phys. Rev. Lett. {\bf 74} (1995) 2768.
\bibitem{Bed95} G. Bed\"urftig, H. Frahm, J. Phys. {\bf A:} Math.
   Gen. {\bf 28} (1995) 4453.
\bibitem{Ram96} P.B. Ramos, M.J. Martins, Nucl. Phys. {\bf B474}
   (1996) 678.
\bibitem{Pfa96} M.P. Pfannm\"uller, H. Frahm, Nucl. Phys. {\bf B479}
   (1996) 575.

\end{thebibliography}
\end{document}